\documentclass[trackchanges]{aastex701}

\usepackage{acronym}
\usepackage{bm}
\usepackage{balance}
\usepackage{graphicx}
\usepackage{amsmath}

\begin{document}

\title{Population of Binary Black Holes Inferred from One Hundred and Fifty Gravitational Wave Signals.}

\author[orcid=0000-0002-1602-4176,sname='V.Tiwari']{Vaibhav Tiwari}
\affiliation{Institute of Gravitational Wave Astronomy, School of Physics and Astronomy, University of Birmingham, Edgbaston}
\email[show]{vaibhavtewari@gmail.com}  

\begin{abstract}

The LIGO–Virgo–KAGRA collaborations have reported \ac{GW} signals from more than 150 \acp{BBH} in the fourth catalog (GWTC-4). Here, we investigate the population properties of these \acp{BBH} using the mixture-model framework Vamana. We present one-dimensional distributions of masses and spins, explore their correlations, and examine their evolution with redshift. These features may reflect astrophysical processes associated with \ac{BBH} formation channels, although most remain poorly constrained. A notable feature is a peak near $10M_\odot$ in the primary mass and $8M_\odot$ in the chirp mass. Additionally, the primary and secondary masses correlate uniquely, producing pronounced chirp-mass peaks around $14M_\odot$ and $27M_\odot$. The three peaks are roughly separated by a factor of two. A simple explanation for such well-placed peaks is a hierarchical merger scenario, in which the first peak dominantly arises from mergers of \acp{BH} of stellar origin, and higher-mass peaks arise from repeated mergers of \acp{BH} from lower-mass peaks. Although most binaries do not exhibit the high spins and characteristic mass ratios expected from hierarchical mergers, those that do are associated with the peaks observed in the chirp mass distribution.

\end{abstract}

\acrodef{GW}[GW]{gravitational wave}
\acrodef{PE}{Parameter Estimation}
\acrodef{MC}{Monte Carlo}
\acrodef{MCMC}{Markov Chain Monte Carlo}
\acrodef{PDF}{Probability Density Function}
\acrodef{NS}{Neutron Star}
\acrodef{BH}{Black Hole}
\acrodef{BBH}[BBH]{Binary Black Hole}
\acrodef{BNS}[BNS]{Binary Neutron Star}
\acrodef{CBC}{Compact Binary Coalesence}
\acrodef{EDF}{empirical distribution function}
\acrodef{CDF}{cumulative distribution function}
\acrodef{SNR}{Signal to Noise Ratio}



\section{Introduction}
With the release of the fourth gravitational-wave catalog, the LIGO–Virgo–KAGRA (LVK) collaborations have more than doubled the number of gravitational-wave signals~\citep{2015CQGra..32g4001L, 2015CQGra..32b4001A, 2021PTEP.2021eA101A, 2025arXiv250818082T, 2025arXiv250818083T}. A key objective of gravitational-wave astronomy is to understand the physical processes that form and merge compact binaries. Often, the approach is to reconcile the anticipated distribution from proposed formation scenarios with the features observed in the compact binary mass, spin, and redshift distributions. Thus, the development of methodologies that accurately infer the population and connect the inferred population to proposed formation scenarios is essential.

Significant progress has been made in developing parametric and non-parametric methods to infer the compact binary population. Parametric methods have employed combinations of power laws, normal distributions, tapering functions, and notch filters to infer the \ac{BBH} mass distribution~\citep{2018ApJ...856..173T, 2019PhRvD.100d3012W, 2019ApJ...882L..24A, 2020ApJ...899L...8F, 2022ApJ...931..108F, 2023arXiv231203973G, 2025PhRvD.111l3046G}. Non- and semi-parametric methods utilising a large number of parameters have revealed additional details in the \ac{BBH} mass distribution~\citep{2021CQGra..38o5007T, PhysRevD.105.123014, 2022MNRAS.509.5454R, 2023ApJ...946...16E, o3b_rnp, 2023MNRAS.524.5844T, 2024ApJ...960...65S, 2024arXiv240403166R, 2024PhRvX..14b1005C, 2024arXiv240616813H}. The mass distribution shows an emerging structure composed of several peaks. The first peak is observed around a chirp mass of 8$M_\odot$ (10$M_\odot$ in the primary mass, which is heavier between the two and, by convention, always greater than the secondary mass)~\citep{2021ApJ...913L..19T, 2021ApJ...917...33L}. The second peak is observed around a chirp mass of 14$M_\odot$~\citep{2021ApJ...913L..19T, o3b_rnp, 2024MNRAS.527..298T}. However, the second peak is not statistically significant in the primary mass~\citep{2023ApJ...955..107F, 2024PhRvX..14b1005C}. A third peak is observed around a chirp mass of 28$M_\odot$ (35$M_\odot$ in primary mass)~\citep{2018ApJ...856..173T, 2019ApJ...882L..24A, 2025arXiv250701086R}.

Significant effort is also underway to investigate the spin distribution of \acp{BBH}~\citep{2022PhRvD.106j3019T, 2022A&A...668L...2V, 2023PhRvD.108j3009G, 2023Univ....9..507P, 2023ApJ...946...50B, 2024ApJ...964L...6A, 2022ApJ...932L..19B, 2024arXiv241102252H, 2025arXiv250106712B, 2025arXiv250204278B}. So far, drawing confident astrophysical implications from spin distributions has been challenging, due in part to a lack of strong signatures in the distribution~\citep{2025arXiv250818083T}, and in part due to large uncertainties in the spin magnitudes predicted by formation scenarios~\citep{2020A&A...635A..97B, 2020A&A...636A.104B, 2022ApJ...933...86Z, 2024MNRAS.534.1868G}. Additional uncertainty arises from processes that can alter spin orientations~\citep{2017PhRvL.119a1101O, 2021PhRvD.103f3032S, 2024arXiv241203461B}, compounded by our limited ability to measure spin components orthogonal to the orbital angular momentum~\citep{2024PhRvD.109j4036M, 2025PhRvD.111b3037H}. Marginal mass and spin distributions together can better constrain the physics of formation channels predicting both distributions~\citep{2020A&A...636A.104B, 2021MNRAS.500.3002B, 2022MNRAS.511.5797M}, and the joint mass-spin distribution is now available from multiple analyses~\citep{2022ApJ...928..155T, 2024arXiv240616844H, 2024arXiv240403166R, 2024A&A...692A..80P, 2025arXiv250602250S}. However, due to the small number of \ac{GW} observations, large model uncertainties, and the variety of proposed formation scenarios~\citep{2022LRR....25....1M}, investigations focusing on mass-dependent spin distributions remain limited. Mass ratios and their correlations with spins can reveal imprints of formation channels~\citep{2017Natur.548..426F, 2021ApJ...922L...5C, 2021ApJ...921L..15G, 2023PhRvD.108j3009G, 2023ApJ...958...13A, 2024MNRAS.531.4725K, 2024PhRvD.109j3006H, 2024arXiv240616844H, 2025PhRvD.111b3037H}, but no channel-specific trend has yet emerged.

Motivated by the two dominant peaks around 10$M_\odot$ and 35$M_\odot$ in the primary mass distribution and the variation of spins with mass, multiple analyses suggest the presence of two dominant sub-populations contributing to \ac{BBH} mergers~\citep{2023arXiv230401288G, 2024arXiv240403166R, 2024PhRvL.133e1401L, 2024arXiv240619044A, 2024arXiv240601679P, 2024arXiv241102252H, 2025arXiv251008231L} (alternatively, see \citealt{2024PhRvD.110f3031H} for a three-sub-population model producing three peaks in the chirp mass distribution, \citealt{2025PhRvD.111b3013M} for hierarchical mergers fitted to primary mass peaks, and \citealt{2025arXiv250522739A} for multiple sub-population fits.). However, the additional peak around 14$M_\odot$ complicates the interpretation.

Investigations for specific scenarios are comparatively straightforward. For example, unique features in the mass and spin distributions are characteristic of a hierarchical merger scenario~\citep{2020ApJ...893...35D, 2024ApJ...966L..16P, 2021MNRAS.507.3362T, 2025PhRvD.111b3013M}. In this scenario, the remnant from a merger may be retained by the environment, forming another binary that merges within a Hubble time~\citep{2021NatAs...5..749G, 2019MNRAS.486.5008A, 2021MNRAS.505..339M, 2023MNRAS.526.4908C}. 

In a specific case, the first-generation~(1G) \acp{BBH} are dynamically formed from \acp{BH} of stellar origin, distributed over a narrow mass range due to a mass-gap pile-up. As a result, the first generation of \acp{BBH} would have comparable masses. The remnants of these 1G mergers constitute second-generation (2G) black holes with masses around twice those of the 1G black holes. A small fraction of remnants retained by the environment can form heavier BBHs. This process can continue, creating higher-generation black holes. In this case, successive peaks in the component mass distribution appear roughly a factor of two apart\footnote{Around 4–6\% of mass is lost to gravitational waves, so the factor is slightly below two.} which would exhibit as distinct peaks in the chirp mass distribution. After the release of GWTC-2, we reported four well-placed peaks in the chirp mass distribution following this multiplicative factor~\citep{2021ApJ...913L..19T}, and additional observations after GWTC-3 retained these peaks~\citep{o3b_rnp, 2022ApJ...928..155T}. While hierarchical mergers provide a simple explanation for these well-placed peaks, two predicted signatures remain largely absent at the population level: the scenario predicts high-spin black holes with dimensionless spins peaking around 0.7, and inter-generation mergers yielding mass ratios of roughly 0.5, 0.25, …, ~\citep{2010CQGra..27k4006L, 2023ApJ...951..129F}. Notably, the high-spin \acp{BBH} we observe do exhibit the expected mass ratios and spin magnitudes. Moreover, their measured masses also coincide with the peaks in the chirp mass distribution -- a correlation that is difficult to explain if the peaks arise from unrelated processes.

In this article, we report the population properties of \acp{BBH} inferred using the mixture-model framework, Vamana. The article is organised as follows: in Section~\ref{sec:method} we briefly discuss our methodology, in Section~\ref{sec:pop} we report the population properties of \acp{BBH}, in Section~\ref{sec:highspin} we discuss the high-spin population, and in Section~\ref{sec:conclude} we summarise our results.
\section{Method}
\label{sec:method}
We use the mixture model framework, Vamana~\citep{2021CQGra..38o5007T}, to infer the population properties of the \acp{BBH}. In the previous version of Vamana, the chirp mass and mass ratio were used as the mass parameters. In the current version, we use the primary and secondary masses instead. In addition to masses, Vamana also infers the aligned spin (the component of spin aligned with the orbital angular momentum) and the redshift evolution of the merger rate, i.e. $R_z = R_0\,(1 + z)^\kappa$. The components in the mixture model are multivariate Gaussians used to infer the mass and spin distributions. They also include a power law to characterise the redshift evolution of the merger rate. Together, these provide a four-dimensional inference about the population. Appendix~\ref{apndx:model} provides a detailed description of the model.

Several independent groups have reported new \ac{BBH} signals after analysing the open data from the LVK~\citep{Venumadhav:2019tad, Nitz:2018imz, Zackay:2019tzo, Venumadhav:2019lyq, Nitz:2020oeq, Nitz:2021uxj, Olsen:2022pin, 2023arXiv231206631W, Nitz:2021zwj, Kumar:2024bfe, Mishra:2024zzs, Mehta:2023zlk, Koloniari:2024kww}. However, since sensitivity estimates are required to accurately infer the population~\citep {2018CQGra..35n5009T}, we use only observations reported by the LVK collaborations~\citep{2019PhRvX...9c1040A, 2021PhRvX..11b1053A, 2024PhRvD.109b2001A, 2023PhRvX..13d1039A, 2025arXiv250818082T}. We select all observations with a false alarm rate of at most one per year and a mean secondary mass greater than 3$M_\odot$. These criteria are satisfied by 153 observations, 69 of which were reported by the LVK in GWTC-3. The estimated parameters for these \acp{BBH} are publicly available~\citep{2023ApJS..267...29A}. Figures presented in Section~\ref{sec:highspin} also use data from the recently released~(and not part of GWTC-4.0) observations GW241110\_124123 and GW241011\_233834~\citep{o4a_special}.
\subsection{Chirp Mass as the Mass Parameter}
\label{subsec:massladder}
We used chirp mass and mass ratio as the two mass parameters in the previous version of Vamana. Chirp mass is the most dominant term that affects the phase evolution of a binary~\citep{Cutler:1994ys}. It is measured with the smallest uncertainty and is the least affected by any biases. The inferred chirp mass distribution is expected to most accurately reflect its true astrophysical counterpart. As a result, chirp mass is sometimes used as the mass parameter. In principle, any pair of mass parameters can be used to infer the underlying mass distribution, and the distributions of the other mass parameters can then be obtained through the appropriate transformations.  In practice, however, population models often lack the flexibility to accurately capture the complex Jacobians involved in these parameter transformations. For example, the chirp mass distribution inferred by applying a parameter transformation to the primary mass and mass ratio may differ from that inferred directly when the chirp mass is used as the mass parameter. 

Despite its accuracy, chirp mass may not be the most appropriate mass parameter, as it can introduce additional features when \acp{BH} with substantially different masses merge at high rates. In the results presented in this article, we have changed the parameters to two-component masses. As a mixture model, Vamana can capture broader variations in secondary mass relative to primary mass. The introduction of a correlation term improves its flexibility for inferring narrow and complex features in the component mass plane. This combined flexibility enables the presented version of Vamana to infer the chirp mass distribution as flexibly as previous versions that inferred it directly. We have demonstrated this in Appendix~\ref{apndx:chirpmass}.

%
\begin{figure*}[t]
\centering
\includegraphics[width=0.89\textwidth]{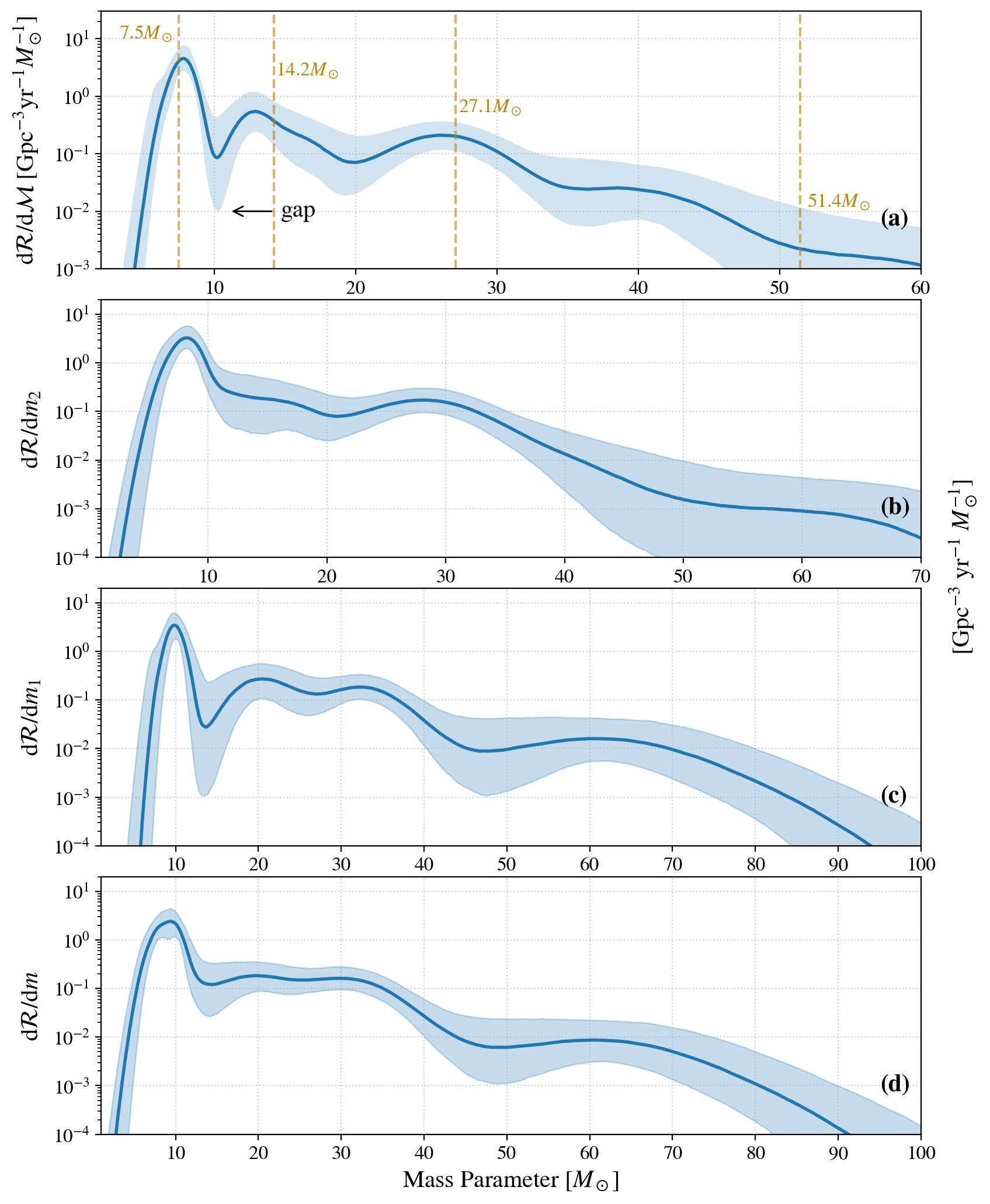}
\caption{The (a) chirp, (b) secondary, (c) primary, and (d) component mass distribution of \acp{BBH}. Solid curves indicate the median, and shaded bands show the 90\% credible intervals for the differential merger rate. All mass parameters show an overdensity at the lower-mass end. The chirp mass distribution exhibits four distinct peaks, with the first three exceeding a 99\% confidence level. The region between the first two peaks is labelled 'gap' due to a lack of observations. Subsequent peaks are separated by approximately a factor of 1.9, shown by brown lines $\mathcal{L}$. The fourth in this set of lines doesn't match with an overdensity, but it has been drawn to mark any features that may arise in future \ac{GW} catalogues}. Other mass parameters show local maxima at comparable locations, but with lower confidence. Primary and secondary masses uniquely correlate to produce the peaks in the chirp mass distribution (see sub-section~\ref{subsec:corrmass}).
\label{fig:mass_distr}
\end{figure*}
\section{Population Properties}
\label{sec:pop}
Vamana infers the two-component masses, the aligned spin components, and the redshift evolution of the merger rate. The aligned spins are assumed to be identically but independently distributed for both \acp{BH}. In this section, we present the inferred distributions marginalised over the modelled or transformed parameters and the correlations between them.
\subsection{Marginalised Distributions in One Dimension}
\begin{figure*}
\centering
\includegraphics[width=0.96\textwidth]{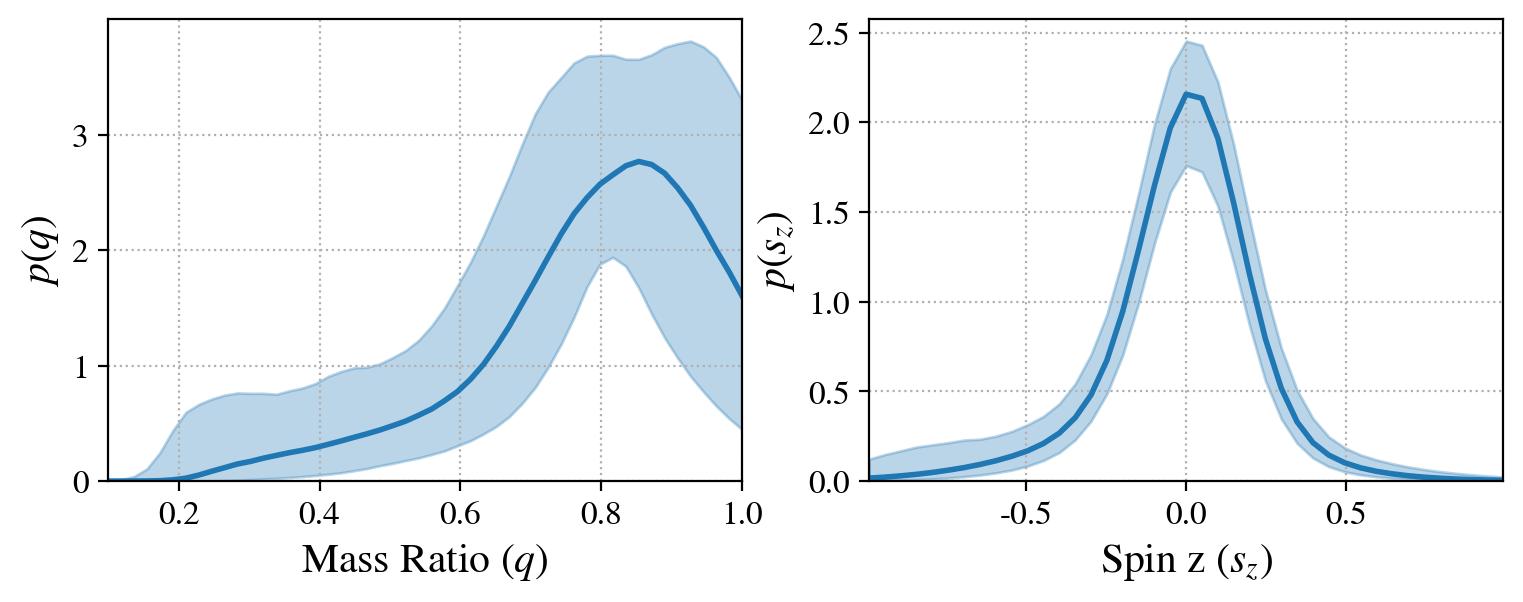}
\caption{Inferred mass-ratio (left) and aligned-spin (right) distributions. Solid lines indicate the median, shaded bands the 90\% credible interval. The mass ratio is mostly uniform from 0.6 to 0.9. Aligned spins are typically small, ranging from -0.30 to 0.37 at 90\% credibility. These figures plot the marginalised distribution. The variation as dependent on various mass parameters is discussed in Section~\ref{subsec:corr}.}
\label{fig:q_and_sz}
\end{figure*}
Figure~\ref{fig:mass_distr} shows the inferred distributions of primary, secondary, component, and chirp masses. These are obtained by marginalising over spins and redshift distributions, yielding the distribution on the component mass plane defined by the primary and secondary masses. Marginalising over the secondary~(primary) mass then yields the distribution of the primary~(secondary) mass. If the designation between the primary and secondary is removed, i.e., $p(m_2, m_1) = p(m_1, m_2)$, one obtains the component mass distribution. Finally, transforming parameters as $(m_1, m_2) \Rightarrow (\mathcal{M}, q)$ produces the chirp mass and mass-ratio distributions.
All distributions show a prominent over-density at the lower end of the mass spectrum, supported by more than 40 observations. The first peak in the chirp mass distribution spans 6–10$M_\odot$. It contributes approximately 66\% to the astrophysical merger rate of \acp{BBH}. However, this fraction is sensitive to the prior on the Gaussian locations for the masses. Currently, we adopt a uniform prior, $p(\mu_{m_1}; \mathrm{or}; \mu_{m_2}) = \mathrm{constant}$. A steeper prior, e.g., $p(\mu_{m_1}, \mu_{m_2}) = 1/(\mu_{m_1}, \mu_{m_2})$, increases this fraction to 78\%. The chirp mass distribution exhibits a local minimum around 10$M_\odot$, where the mean density drops by a factor of 28 from 7.2$M_\odot$ to 10.2$M_\odot$, and then rises by a factor of 4 at 12.9$M_\odot$. This `gap' corresponds to a lack of observations with mean chirp masses between 8.9 and 11.7$M_\odot$ and is discussed in Section~\ref{sec:highspin} in the context of a hierarchical merger scenario. Using methodologies from \citet{2024MNRAS.527..298T} and \citet{2025arXiv250715697T}, we estimate over 98\% confidence in the peak around 14$M_\odot$. It is important to note that the presented model does not directly infer the chirp mass distribution and, therefore, the second peak is not a consequence of mixture models' tendency to infer peaks and gaps. We have discussed this further in Appendix~\ref{apndx:chirpmass} and ~\ref{apndx:confpeak2}. We have not estimated confidence in the presence of a peak at chirp mass 26$M_\odot$, but given the modulation in the median distribution and the narrow credible bands, confidence in this peak is expected to be high~(see \citet{2025arXiv250701086R} for a detailed study of this peak).
Other mass parameters exhibit modulations at comparable locations, but with limited confidence. For distributions where component masses are uncorrelated, a peak in the component mass distribution typically produces a peak in the chirp mass at a slightly smaller value (by a factor of $2^{0.2}$) and in the primary mass at a value roughly 1.2 times larger. The presence of a confident peak in the chirp mass distribution, without a comparable peak in the primary or secondary masses, implies a unique correlation between the component masses, discussed further in sub-section~\ref{subsec:corrmass}.

Figure~\ref{fig:q_and_sz} shows the inferred mass-ratio and aligned spin distributions. In previous reports, the mass-ratio distribution was found to peak at unity~\citep{2022ApJ...928..155T}, an artefact of fitting a power law to infer the distribution. Modelling component masses with Gaussians allows for more flexible inference. The inferred distribution is uniform, mainly between 0.6 and 0.9, with a wide credible interval due to limited precision in mass-ratio measurements. This suggests no strong evidence for preferential pairing of comparable-mass binaries. The aligned spin distribution is symmetric, peaking around 0.05, with most observations consistent with small aligned spins. Priors on model hyperparameters have a noticeable impact on the inferred mass ratio and aligned spin distribution~\citep{baird-2013, 2017PhRvL.119y1103V, 2018ApJ...868..140T} as these parameters are strongly correlated and are measured with significant uncertainty. A prior that favours smaller aligned spin values results in a larger inferred mass ratio, and vice versa.

\begin{figure}
\centering
\includegraphics[width=0.48\textwidth]{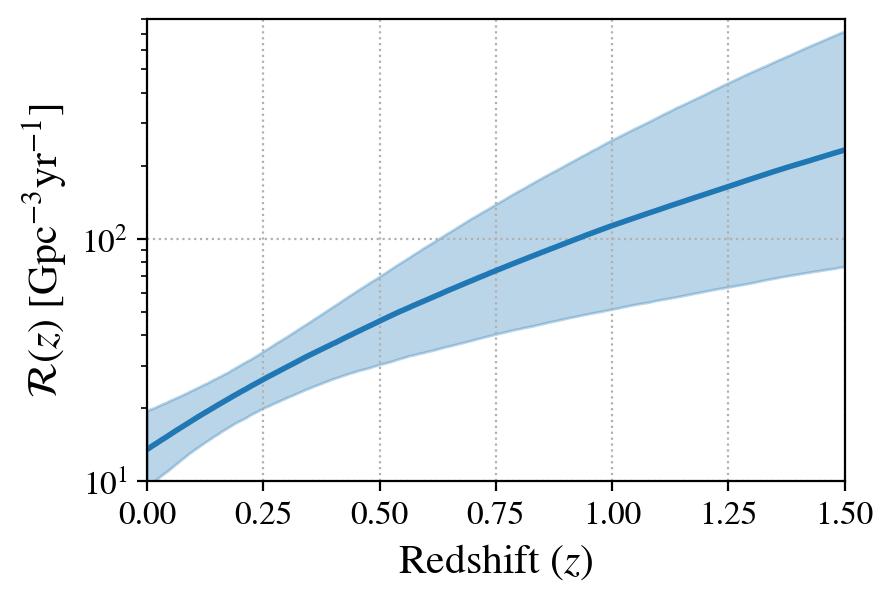}
\caption{Local merger rate (left) and redshift evolution (right). Solid lines indicate the mean, shaded bands the 90\% credible interval. The local merger rate is $R_0 = 14.0^{+4.8}_{-5.9}\mathrm{Gpc}^{-3},\mathrm{yr}^{-1}$. Redshift evolution is confidently positive.}
\label{fig:Rz}
\end{figure}

Figure~\ref{fig:Rz} shows the redshift evolution of the merger rate. Models assumes a power-law evolution of the merger rate, $R(z) = R_0 (1 + z) ^\kappa$~\citep{2018ApJ...863L..41F}, Where $R_0$ is the merger rate at redshift zero. Merger rate is estimated to be, $R_0 = 14.0^{+4.8}_{-5.9}\mathrm{Gpc}^3\,\mathrm{yr}^{-1}$. The right side of Figure~\ref{fig:Rz} shows the evolution of the merger rate with the redshift. The evolution is confidently positive with $\kappa$ ranging from 0.4--4.7 at 90\% credibility. Priors make a noticeable impact on the inferred values of $\kappa$ and the merger rate. A prior that favours populations with larger sensitive volumes results in smaller merger rates, e.g., a prior on $\kappa$ that extends to large positive values.
\subsection{Correlation between parameters}
\label{subsec:corr}
We now examine correlations between the modelled or transformed parameters. Instead of computing posterior predictive distributions on a multidimensional grid (with Jacobian corrections for transformations), we draw a large number of samples for component masses, aligned spins, and redshift. All figures in this sub-section, except the differential merger rate on the component mass plane, are generated by histogramming these samples.

\subsubsection{Correlation between mass parameters}
\label{subsec:corrmass}
\begin{figure*}
\includegraphics[width=0.96\textwidth]{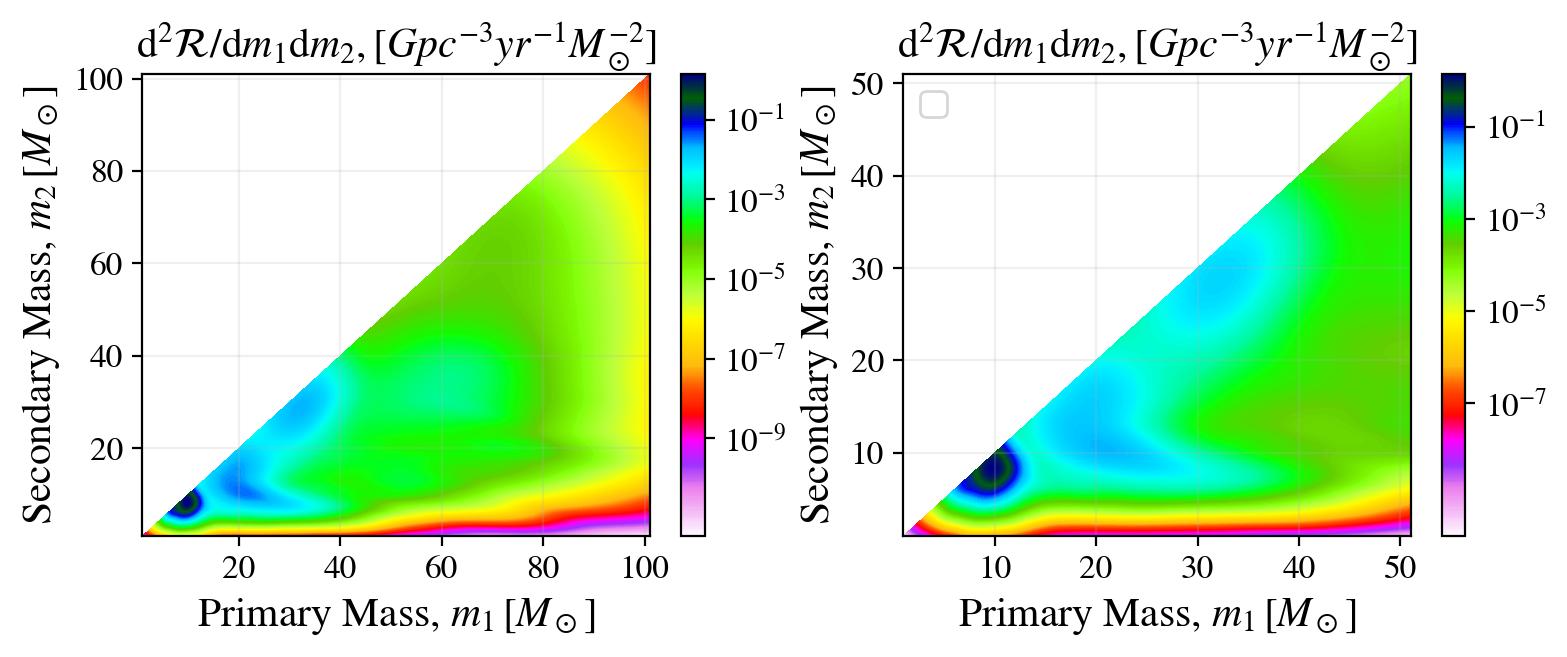}
\centering
\caption{Differential merger rate on the component mass plane~(right plot is lower-left quadrant of the left). Three over-dense regions correspond to peaks in the chirp mass distribution. The dashed curve indicates a constant chirp-mass track, $\mathcal{M} = 13.9M_\odot$, which requires the component masses to be uniquely correlated. The over-density around 60--40$M_\odot$ has been interpreted as due to the presence of a gap in the secondary mass caused by pair-instability supernova~\citep{1964ApJS....9..201F, 2017ApJ...836..244W} and lack of it in the primary mass due to 2G--1G hierarchical mergers, where masses of 1G BHs in this interpretation extends up to around~45$M_\odot$~\citep{2025PhRvD.112f3040A, 2025arXiv250904151T, 2025arXiv250904637A, 2025arXiv250909123A, 2025arXiv250909876G}.}
\label{fig:corr_m1m2}
\end{figure*}

\begin{figure*}
\centering
\includegraphics[width=0.96\textwidth]{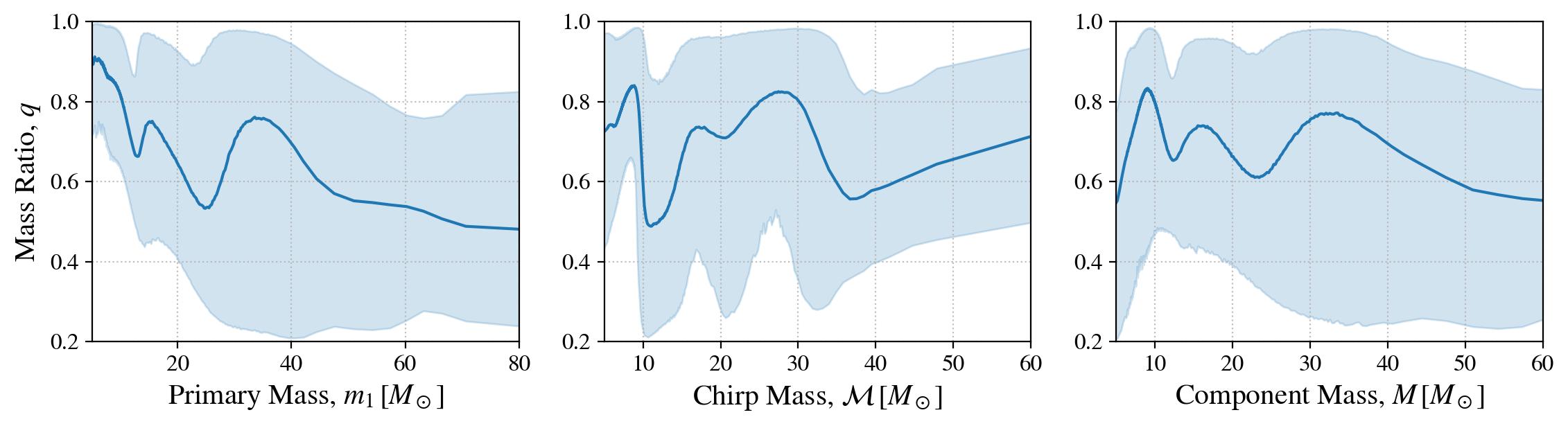}
\caption{Variation of mass ratio with mass parameters. The primary/total mass combines uniquely with the mass ratio to produce the second chirp mass peak at 14$M_\odot$. The jagged structure arises from a change in the sign of the correlation around $m_1 = 19M_\odot$. In contrast, the third peak mostly consists of binaries with comparable masses.}
\label{fig:corr_massq}
\end{figure*}

Figure~\ref{fig:corr_m1m2} shows the differential merger rate on the component mass plane, revealing three prominent over-dense regions corresponding to peaks in the chirp mass distribution. The region associated with the 14$M_\odot$ chirp mass peak exhibits a unique correlation between primary and secondary masses: it produces a prominent chirp mass peak without producing a comparably strong peak in the individual component masses (see Figure~\ref{fig:mass_distr}). This is also evident in Figure~\ref{fig:corr_massq}. The mass ratio varies, forming a jagged curve for both the primary and total mass. The correlation changes sign around a primary mass value of 19$M_\odot$. This feature is more prominent in the primary and almost absent in the chirp mass. We have previously reported this feature for observations in GWTC-3~\citep{2024MNRAS.527..298T, 2025arXiv250715697T}. Chirp mass is the most accurately measured parameter. Although it is not a parameter directly relevant to astrophysical processes, the presence of a strong feature in the chirp mass requires investigation.

The effect of this correlation is reflected in the distribution of mass ratios. Mass ratios for the second peak significantly contrast with those for the third peak located around the chirp~(primary) mass value 27$M_\odot$~(35$M_\odot$). By making random draws for mass ratio corresponding to the second peak~(11.0 < $M_\odot$ < 19.0) and the third peak (19.0 < $M_\odot$ < 32.0), we find the third peak has a larger mass ratio for 85\% of the draws.

\subsubsection{Correlation between mass ratio and aligned spin}
Previous studies report correlations between effective spin and mass ratio~\citep{2021ApJ...922L...5C, 2023ApJ...958...13A}. Using the updated Vamana model, we continue to find an essentially flat relation between aligned spin and mass ratio, with slight broadening at lower mass ratios. Random draws indicate that binaries with $q \in [0.1,0.5]$ have larger aligned spins than those with $q \in [0.5,1.0]$ in 60\% of draws.
\begin{figure}
\centering
\includegraphics[width=0.48\textwidth]{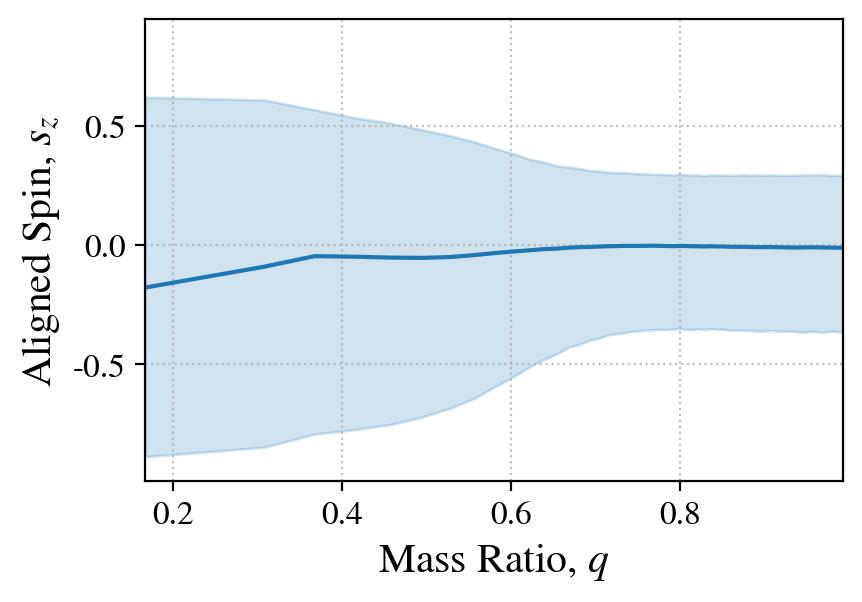}
\vspace{-5mm}
\caption{Variation of aligned spins with mass ratio. The distribution is mostly flat, showing no strong correlation.}
\label{fig:corr_qsz}
\end{figure}

\subsubsection{Correlation between mass parameters and aligned spin}
\begin{figure*}
\centering
\includegraphics[width=0.96\textwidth]{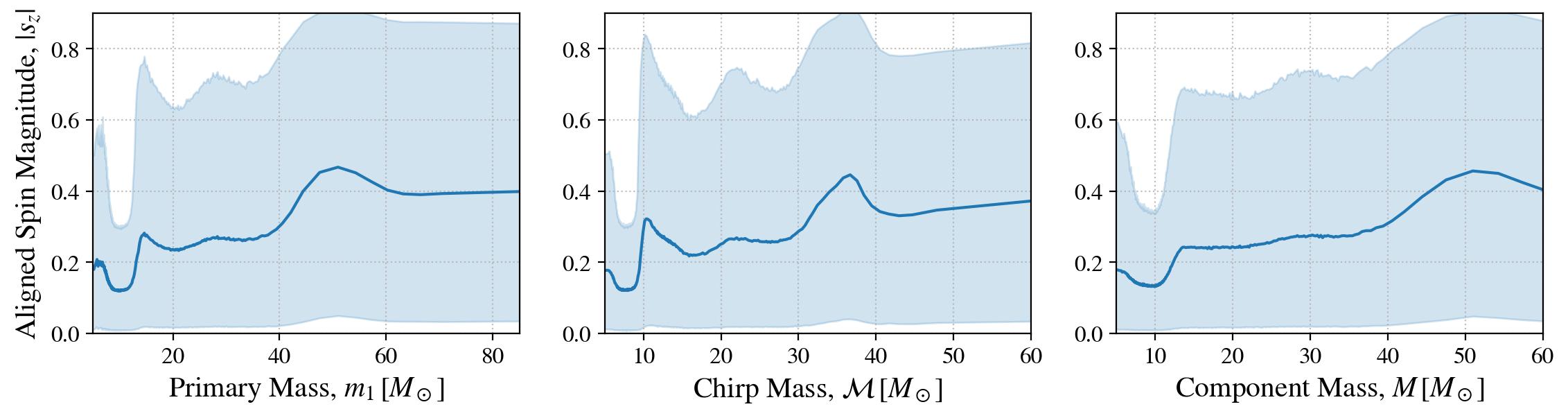}
\caption{Aligned spin as a function of mass parameters. The solid line represents the median, and the blue band indicates the 90\% credible interval. Heavier binaries tend to have larger aligned spins.}
\label{fig:corr_masssz}
\end{figure*}

Figure~\ref{fig:corr_masssz} shows that aligned spins are generally low across most of the mass range, but increase for heavier binaries. For chirp masses below 35$M_\odot$, the 90\% credible interval for aligned spins is 0.05–0.25, rising to 0.1–0.4 above 35$M_\odot$. Random draws indicate heavier binaries have larger aligned spins in 80\% of cases. This trend, along with two dominant primary-mass peaks, motivates the use of multi-population models for \ac{BBH} observations.

\subsubsection{Redshift evolution of the population}
\begin{figure*}
\centering
\includegraphics[width=0.96\textwidth]{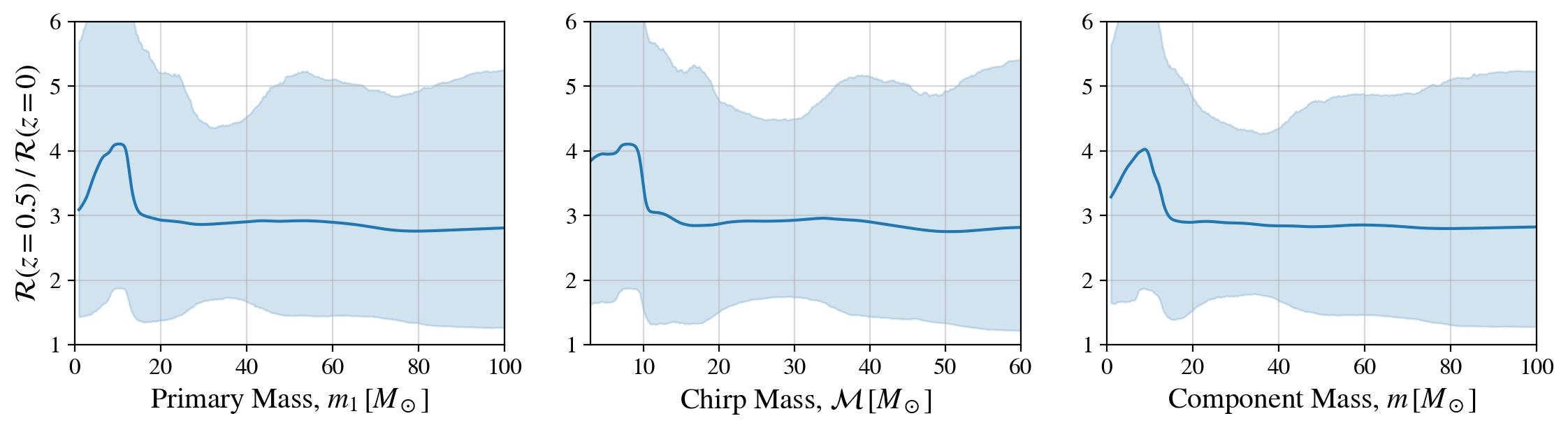}
\caption{Merger rate increase at $z=0.5$ relative to the local universe. The distribution is mostly flat, with no strong evidence for mass-dependent evolution.}
\label{fig:rate_with_mass_evol}
\end{figure*}

Vamana models the redshift evolution of the merger rate as a power law, allowing components in different regions of parameter space to evolve differently. Figure~\ref{fig:rate_with_mass_evol} shows the ratio of merger rates at $z=0.5$ and $z=0$ as a function of mass. Previously, the evolution appeared shallower for the heaviest binaries ($\mathcal{M} > 35 M_\odot$) at 90\% credibility~\citep{2022ApJ...928..155T}, but with additional observations, this confidence drops to 70\%. The luminosity distance, degenerate with the inclination angle, is measured with a large uncertainty, limiting the precision with which population evolution can be measured.
%
\section{Hierarchical Mergers}
\label{sec:highspin}
Observation of peaks in the component mass distribution, with their relative locations bearing a constant factor, is characteristic of a hierarchical merger scenario. In this scenario, the first peak corresponds to the first-generation (1G) mergers between \acp{BH} of stellar origin, and the higher-mass peaks are due to repeated mergers of \acp{BH} remnants from the lower-mass peaks. The relative location of peaks is expected to scale by a factor of around 1.9, accounting for the doubling of masses and an approximately 5\% mass loss due to the emission of \acp{GW}~\citep{2012ApJ...758...63B}. The locations of the peaks imply mass ratios of roughly 1:1, 1:0.5, 1:0.25, and other combinations, resulting from intra- and intergenerational mergers. In addition, these binaries often have high spins~\citep{2010CQGra..27k4006L}\footnote{If the spin magnitude is narrowly distributed around 0.7 and orientations are isotropic, the aligned spin distribution is approximately uniform, extending from -0.7 to 0.7.}.
\subsection{High-Spin Binary Black Holes}
\begin{figure*} 
\centering
\includegraphics[width=0.96\textwidth]{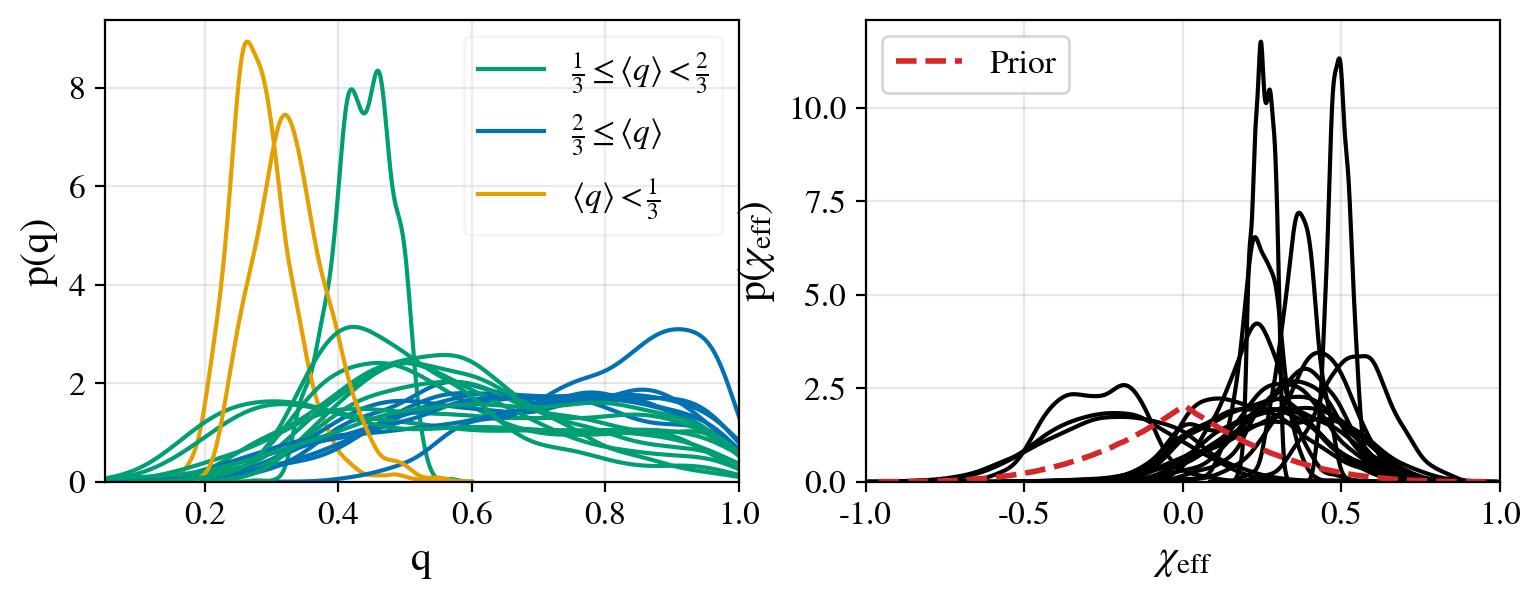} 
\caption{The mass ratio~(left) and effective spin~(right) distribution of \acp{BBH} with $|\langle\chi_\mathrm{eff}\rangle| > 0.2$. Parameters are estimated using a standard prior~\citep{2019ApJS..241...27A}. The prior is approximately uniform for the mass ratio. Combined with the priors on spins, which are assumed to have uniform magnitude and isotropic orientation, the corresponding prior on $\chi_\mathrm{eff}$ is shown in the figure as a dashed red line. We use \ac{BH} masses measured from these binaries to infer the \ac{BH} mass distribution shown in Figure~\ref{fig:hispin_pm}.} \label{fig:raw_q_and_comp} 
\end{figure*}

Although a majority of the population does not exhibit the expected characteristics from the hierarchical merger scenario, binaries with high spins do. Moreover, these high-spin binaries exhibit mass ratios and aligned spins consistent with expectations from hierarchical mergers. Notably, the components of most of the high-spin binaries are associated with the peaks observed in the chirp mass distribution. The first peak in chirp mass is located around 7.5$M_\odot$. The component mass for a comparable-mass binary corresponding to this value is $7.5\times 2^{0.2} = 8.5M_\odot$.
\footnote{For a binary with component masses $m_1$ and $m_2$, the chirp mass is given by
$\mathcal{M} = (m_1 m_2)^{0.6}/(m_1 + m_2)^{0.2}$.
For $m_1=m_2$, the scaling factor is $2^{0.2}$.}
. Consequently, the \acp{BH} involved in intra- and inter-generation mergers will have masses distributed around $8.5M_\odot\times 1.9=16.2M_\odot$ and $8.5M_\odot\times 1.9^2=30.7M_\odot$. We previously noted that, in addition to high spins, the observation GW190412~\citep{2020PhRvD.102d3015A}, with mean masses of 8.3–30.1$M_\odot$, is consistent with a 1G–3G merger~\citep{2021ApJ...913L..19T}. GW151226~\citep{2016PhRvL.116x1103A} and the recently announced GW241011\_23383 and GW241110\_124123 have masses consistent with 1G–2G mergers. Observation GW230624\_113103 has masses consistent with a 2G–3G merger~\citep{2025arXiv250818082T}. Most of the \acp{BH} masses for high-spin binaries are close to the values we calculated.
To demonstrate this, we select high-spin binaries and infer the mass distribution of their components. \acp{GW} provide limited information on the spins of individual \acp{BH}~\citep{2016PhRvD..93h4042P, 2017PhRvD..95f4053V}. An effective spin parameter is measured substantially more accurately and is defined as~\citep{2001PhRvD..64l4013D, 2008PhRvD..78d4021R, 2011PhRvL.106x1101A}:
\begin{equation}
\chi_\mathrm{eff} = \frac{m_1\,s_\mathrm{1z}\,+\,m_2\,s_\mathrm{2z}}{m_1 + m_2},
\end{equation}
for the component masses, $m_1$ and $m_2$, and components of the spins aligned with the orbital angular momentum, $s_\mathrm{1z}$ and $s_\mathrm{2z}$. We select all the \acp{BBH} with mean effective spin magnitude greater than 0.2, i.e. $|\langle\chi_\mathrm{eff}\rangle| > 0.2$. This is not the most robust criterion for choosing high-spin \acp{BBH}, as it cannot discriminate whether one or both \acp{BH} have high spins or account for the presence of in-plane spin. But, overall, the spins of \acp{BH} in the selected binaries are expected to be substantially higher. The mass ratio and spin distribution of the chosen binaries are shown in Figure~\ref{fig:raw_q_and_comp}.

\begin{figure} 
\centering
\includegraphics[width=0.96\textwidth]{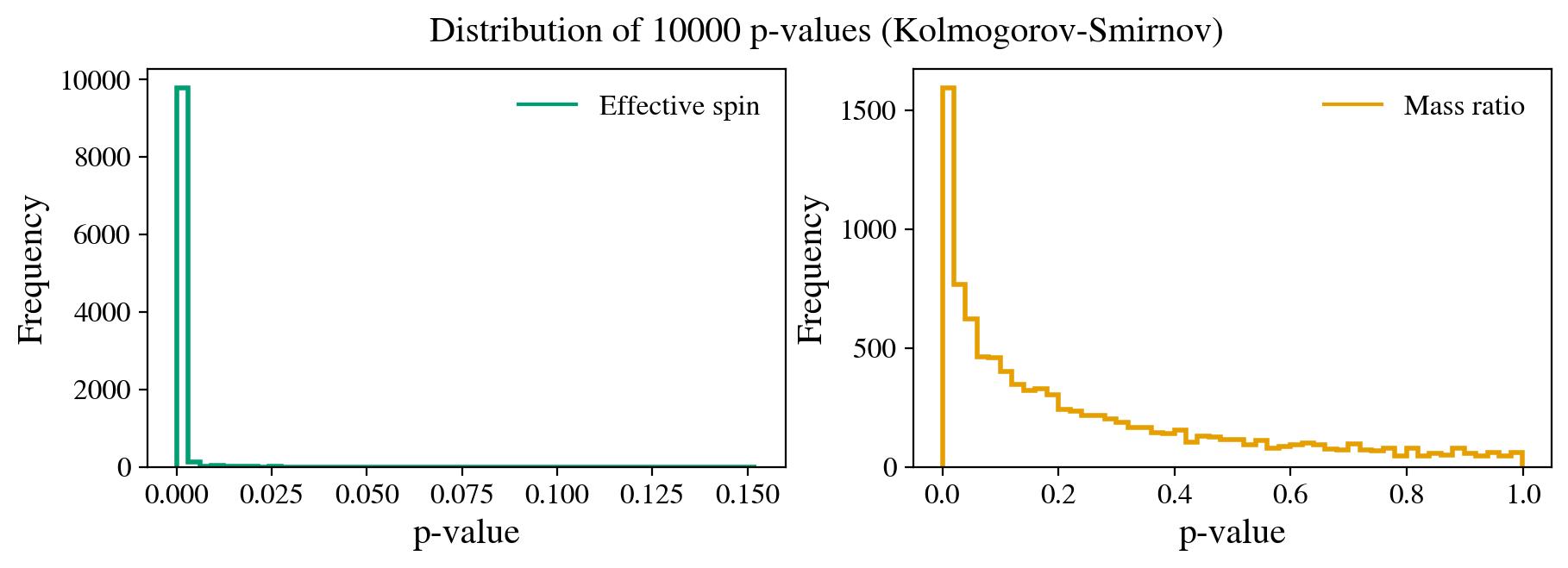} 
\caption{The KS p-values between possible values of mass ratio and effective spins between the high-spin \acp{BBH} used in estimating \ac{BH} mass distribution shown in Figure~\ref{fig:hispin_pm} and \acp{BBH} used in estimating the chirp mass distribution shown in Figure~\ref{fig:mass_distr}. To calculate KS p-values, we randomly selected one sample per observation and computed the KS p-value for that selection. The histogram shown was created from multiple p-values calculated across various selections. The p-values from samples drawn from the same population are expected to be uniformly distributed. However, the histogram's p-values are significantly skewed towards small values. The estimated parameters use a standard prior. In practice, these estimates should be re-weighted to population-informed priors. We don't expect that would meaningfully impact our conclusion.} 
\label{fig:hispin_ks} 
\end{figure}

\begin{center} 
\begin{figure*} 
\centering
\includegraphics[width=0.96\textwidth]{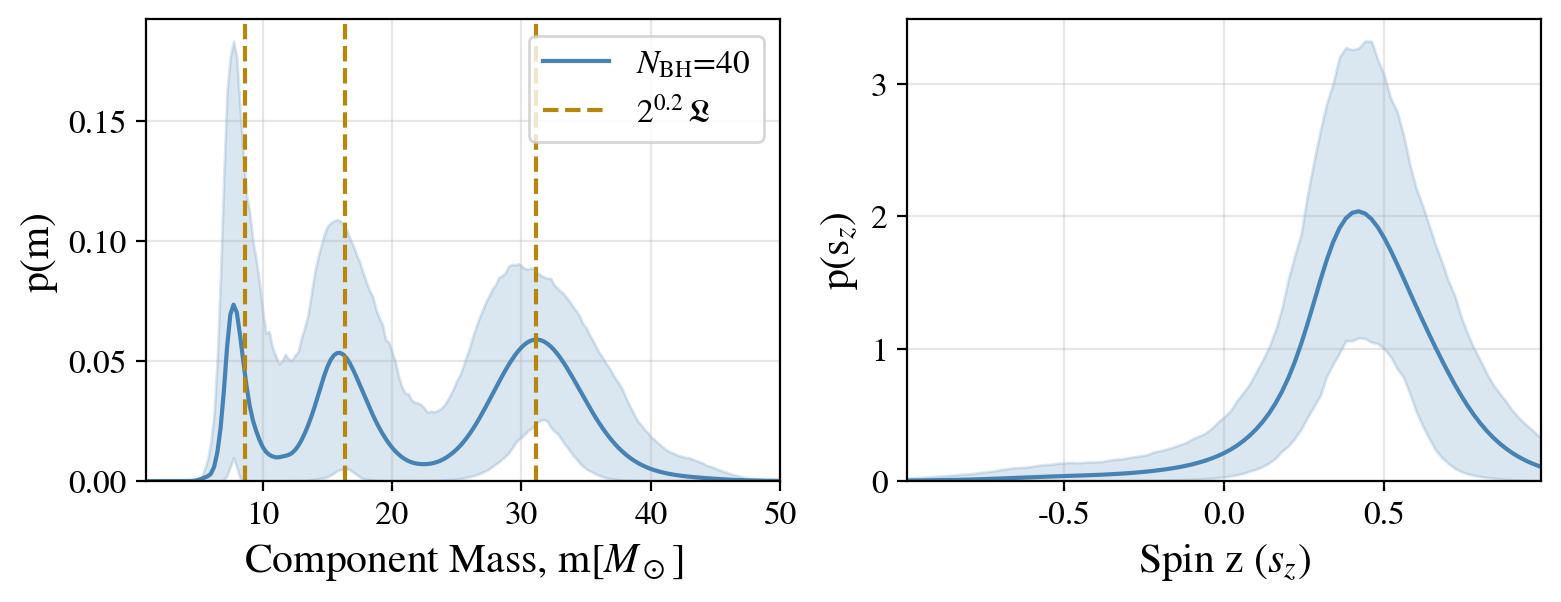} 
\caption{The \ac{BH} mass~(left) and aligned spin~(right) distribution inferred from \acp{BBH} with $|\langle\chi_\mathrm{eff}\rangle| > 0.2$. After accounting for the $2^{0.2}$ scaling factor required to map the chirp mass to the component masses of a comparable-mass binary, the peaks in this figure and in Figure~\ref{fig:mass_distr} match the brown lines well.} \label{fig:hispin_pm} 
\end{figure*} \end{center}

These observations have mass ratios clustered around 1.0 and 0.5. GW190412 has a mean mass ratio of around 0.25. The high-spin binaries constitute a population substantially different from that of \acp{BBH}, as shown in Figure~\ref{fig:hispin_ks}. The spins of the two distributions differ significantly, as expected given our choice. In addition, the two populations also differ in mass ratio as demonstrated in Figure~\ref{fig:hispin_ks}.
Using individual mass and aligned-spin measurements from the selected binaries, we infer the distributions of \ac{BH} mass and aligned spin. The methodology details are provided in Appendix~\ref{apndx:hispin}. These inferences are shown in Figure~\ref{fig:hispin_pm}. The \ac{BH} masses cluster around specific mass values. The brown lines compare the peaks in Figure~\ref{fig:hispin_pm} with the peaks observed in the chirp mass distribution shown in Figure~\ref{fig:mass_distr}. We note that the peaks in the component mass distribution for the high-spin population are complementary to the peaks observed in the chirp mass distribution of the full population.

The inferred distribution has not been corrected for the selection effect. The relative height of the peaks will change after this correction. The redshift evolution has been fixed to be uniform in comoving volume. Choosing a steeper evolution shifts the peak to lower masses. This change is noticeable but small, and it affects heavier mass peaks more. As noted in Appendix~\ref{apndx:hispin}, the pairing probability of \acp{BH} has not been modelled; its inclusion will have a noticeable effect on this structure. Overall, confidence in the structure is low, but it suggests that the component masses of high-spin binaries cluster around specific values and are complementary to the peaks in the chirp mass distribution. This structure has become increasingly pronounced with the increase in the size of the \ac{GW} catalog. Each observation run has accumulated an increasing number of high-spin \acp{BBH} whose components are consistent with the peaks in the chirp mass distribution.
%
%
\section{Conclusion}
\label{sec:conclude}
In this article, we have reported the population properties of binary black holes following the fourth LVK gravitational-wave catalog, which roughly doubles the number of observed events. One-dimensional distributions of mass, mass ratio, and spin, along with their correlations and the redshift evolution of the merger rate, have been presented. While most features remain tentative and require more observations to establish robust confidence, the three peaks in the chirp mass distribution are notable. These peaks are not mirrored at comparable confidence in the primary or secondary mass distributions, reflecting a unique correlation between the component masses. Their locations, spaced roughly by a factor of two, can be simply explained in the context of hierarchical mergers.

While the overall paucity of high spins challenges a naive hierarchical picture, the component masses combine to produce a markedly pronounced structure in the chirp-mass distribution. The chirp mass, being the most precisely measured mass parameter, does not directly map onto formation physics; conversely, spin inferences are less precise and influenced by modelling, priors, and selection effects, potentially underrepresenting a high-spin population. Notably, the subset of high-spin binaries exhibits component masses that align with the peaks of the chirp mass distribution, providing direct evidence for hierarchical scenarios.

Taken together, these observations raise the question of whether hierarchical mergers may dominate the shaping of the binary black hole mass distribution, rather than merely making a minor contribution. Even if spins remain difficult to measure, the mass distribution alone carries signatures of hierarchical processes. The aspects detailed here indicate that hierarchical mergers help shape the binary black hole mass distribution, even at lower masses. Until recently, their contribution at lower masses was not considered important, thus marking a potential shift in our understanding of the formation scenario of binary black holes. With the remaining fourth observation run expected to roughly double the current catalog, these signatures can be tested more rigorously, offering an opportunity to refine our understanding of binary black hole formation channels.

\begin{acknowledgments}
The author thanks Alberto Vecchio, Bernard Schutz, and Fabio Antonini for helpful discussions and Debatri Chattopadhyay for reviewing the manuscript. 
This material is based upon work supported by NSF's LIGO Laboratory which is a major facility fully funded by the National Science Foundation. The author gratefully acknowledges computing resources provided by the LIGO laboratory, supported by the National Science Foundation grants, PHY-0757058 and PHY-0823459, and computing resources provided by Cardiff University, funded by Science and Technology Facilities Council grants, STFC grants ST/I006285/1 and ST/V005618/1.
\end{acknowledgments}

%
\facilities{LIGO, Virgo, KAGRA}


\software{\texttt{NumPy}~\citep{harris2020array}, \texttt{SciPy}~\citep{2020SciPy-NMeth}, \texttt{matplotlib}~\citep{Hunter:2007},  and \texttt{ASTROPY}~\citep{2022ApJ...935..167A}. 
          }


\appendix

\section{Model Details}
\label{apndx:model}
We utilise 10 components in the mixture model. The functional form of Vamana is described in Equation~\ref{eq:model1}. Hyperparameters and their priors are listed in Table~\ref{tab:Lambda}. Vamana infers the joint distribution of primary mass, secondary mass, aligned spin, and the redshift evolution of the merger rate. Components are multivariate normals combined with a power law. The cross-terms for the covariance matrix between different parameters are set to zero, except for the covariance between the primary and secondary mass. This enables the mixture model to accurately infer the chirp-mass distribution. 
The \ac{BBH} population is inferred from observations with a false-alarm rate of at most once per year. We only used observations reported by the LVK collaborations~\citep{2019PhRvX...9c1040A, 2021PhRvX..11b1053A, 2024PhRvD.109b2001A, 2023PhRvX..13d1039A}. Only observations with a mean chirp mass greater than 5$M_\odot$ are used. We excluded GW190814 and any observation consistent with a binary neutron star and a neutron star-black hole binary from our investigations. The total number of observations chosen is 153. The binary parameters are estimated in the detector frame; we assume the Planck15 cosmology \citep{2016A&A...594A..13P} to convert to source-frame quantities. 
\begin{equation}
\begin{split}
p(m_1,m_2,\chi_1,\chi_2|\Lambda) =
&\sum_{i=1}^{N} w_i \, 
\mathcal{N}(m_1,m_2|\mu_i^{m_1},\sigma_i^{m_1},\mu_i^{m_2},\sigma_i^{m_2},C_i^{m_1 m_2})\\
&\qquad \times \phi(\chi_1|\mu_i^\chi,\sigma_i^\chi)\,\phi(\chi_2|\mu_i^\chi,\sigma_i^\chi).
\end{split}
\label{eq:model1}
\end{equation}
\begin{deluxetable*}{lccc}
\label{tab:hyperparams}
\tablecaption{Hyperparameters of the model used to infer the BBH population. U stands for Uniform, and UL for Uniform-in-log. $N$ is the number of components. Analysis imposes constraints, $\mu_i^{m_1} \geq \mu_i^{m_2}$ and $\mu_i^{m_2}/\mu_i^{m_2} \geq 0.2$ on the location of Gaussians. The merger rate of the population is modelled using a power law, $\mathcal{R} \propto (1 + z)^{\kappa_p}$.\label{tab:Lambda}}
\tablewidth{0pt}
\tablehead{
\colhead{$\Lambda$} & \colhead{Description / Modeled Parameter} & \colhead{Prior} & \colhead{Range}
}
\startdata
$w_i$ & Mixing weights, $w$ & Dirichlet($\bm{\alpha}$), $\alpha_{1\cdots N}= 1/N$ & 0--1 \\
$\kappa_\mathrm{pop}$ & Populations's merger rate evolution index, $z$ & U / UL & $|\kappa_\mathrm{pop}| < 1.0$ / $1.0 < |\kappa_\mathrm{pop}| < 5.0$ \\
$\kappa_\mathrm{comp}$ & Components's merger rate evolution index, $z$ & U & $|\kappa_\mathrm{pop} - \kappa_\mathrm{comp}| < 1.5$ \\
$\mu^\chi_i$ & Gaussian's location, $s_z$ & U / UL & $|\mu^\chi_i| < 0.5$ / $0.5 < |\mu^\chi_i| < 0.9$ \\
$\sigma^\chi_i$ & Gaussian's scale, $s_z$ & U & 0.5 / $\sqrt{N}$--1.5 / $\sqrt{N}$ \\
$\mu_i^{m_1}$ & Gaussian's location, $m_1$ & U & 6$M_\odot$--75$M_\odot$ \\
$\sigma_i^{m_1}$ & Gaussian's scale, $m_1$ & U & 0.06\,$\mu_i^{m_1}$ -- 0.18\,$\mu_i^{m_1}$ \\
$\mu_i^{m_2}$ & Gaussian's location, $m_2$ & U & 6$M_\odot$--75$M_\odot$ \\
$\sigma_i^{m_2}$ & Gaussian's scale, $m_2$ & U & 0.06\,$\mu_i^{m_2}$ -- 0.18\,$\mu_i^{m_2}$ \\
$C_i^{m_1\text{--}m_2}$ & Covariance, $m_1$--$m_2$ & U & -0.75\,$\sigma_i^{m_1}\,\sigma_i^{m_2}$ -- 0.75\,$\sigma_i^{m_1}\,\sigma_i^{m_2}$ \\
\enddata
\end{deluxetable*}

We have chosen a uniform prior on the locations of the Gaussians. Using a prior, such as a decaying power law, that favours low masses would affect the inference accordingly. However, we don't expect this choice to affect the features inferred in the mass distributions. The corresponding scale is proportional to their location. This choice is motivated by uncertainty in how mass is measured across different mass values. Vamana employs a large number of hyperparameters and usually requires a separate analysis to estimate confidence in the features inferred from the modelled parameters. 

We have not introduced a correlation term between the masses and spins. We have verified that introducing such a term has a small impact. Being a mixture model, Vamana can infer broad variations of spins with respect to other modelled parameters. Introducing a correlation term to infer narrow features is not beneficial due to the limited information available from spin measurements.

Unless the number of components is too small or too large, they play a small role in the analysis. This is because the scales of the Gaussians are adjusted to effectively provide a similar kernel bandwidth. This leaves only priors on the scales that need to be tuned. We have used 10 components in this analysis. The inferences only change slightly when the number of components is increased or decreased. 

\section{Inferences on Chirp mass distribution}
\label{apndx:chirpmass}

\begin{center} 
\begin{figure*} 
\centering
\includegraphics[width=0.96\textwidth]{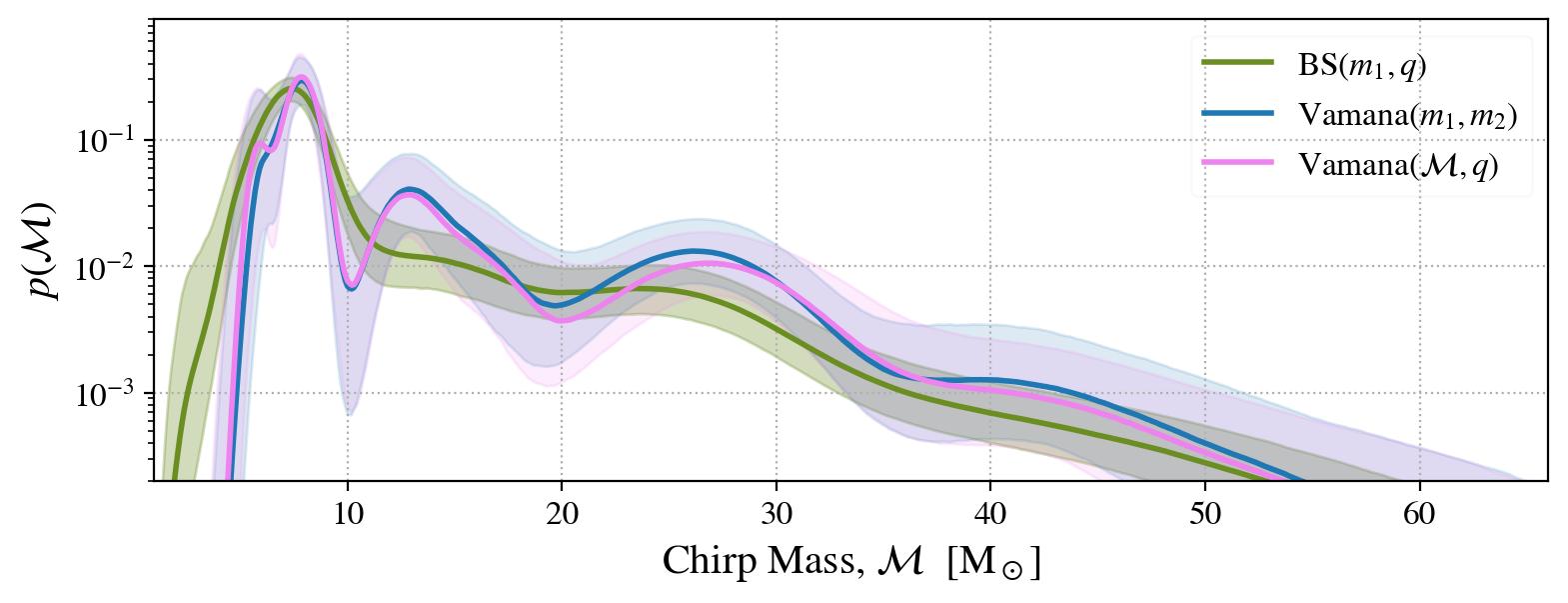} 
\caption{The chirp mass distribution inferred using previous and current versions of Vamana. The figure also shows the chirp mass inferred by the BSpline model. This model assumes a uniform mass-ratio distribution across all masses. The solid line indicates the median, and the band is the 90\% credible interval. The brackets in the legend enclose the mass parameters directly inferred by the corresponding model. Only Vamana($\mathcal{M}, q$) infers the chirp mass directly. For the remaining two models, the chirp mass distribution is obtained by performing a parameter transformation on the inferred mass parameters. The primary mass inferred by model Vamana($m_1, m_2$) is shown in Figure~\ref{fig:mass_distr} and for model BS($m_1, q$) is reported in\citealt{2025arXiv250818083T}.}
\label{fig:compare_mchirp} 
\end{figure*} 
\end{center}

In this section, we compare the chirp mass distribution inferred using the version presented in this article with that inferred by a previous version of Vamana, which directly models the chirp mass. We have omitted details of this earlier version, but they are present elsewhere~\citep{2021CQGra..38o5007T}. Figure~\ref{fig:compare_mchirp} shows the inference on chirp mass by applying a parameter transformation on the distribution inferred on the component mass plane and inference made by directly modelling the chirp mass. The priors for the two models are not equivalent because the modelled parameters differ. Despite that, the two inferences are closely aligned in terms of the median and credible intervals. The presented version of Vamana infers the component masses while retaining the flexibility to model the chirp-mass distribution. As a result, we no longer use the chirp mass as a mass parameter. This figure also includes the chirp mass distribution inferred by the BSpline model~\citep{2023ApJ...946...16E, 2025arXiv250818083T}. This model uses the primary mass and mass ratio as the two mass parameters; we apply a Jacobian to obtain the chirp-mass distribution. 

\section{Confidence in the second chirp mass peak}
\label{apndx:confpeak2}

The median chirp mass distribution in Figure~\ref{fig:mass_distr} shows a feature that comprises a dip followed by a peak. Following~\citealt{2025arXiv250715697T}, we calculate the confidence in this feature. We first infer the chirp mass distribution using the mixture model Vamana~($\mathbb{M}$) and a flavour of it that avoids inferring a feature in the chirp mass range 9.5--20$M_\odot$ by modelling the distribution using a powerlaw~($\mathbb{M}_{pl}$). These models are described in~\citealt{2025arXiv250715697T}. 

We examine whether the feature in the chirp mass distribution can be confidently explained as just a Poisson fluctuation from a featureless distribution. We use two statistics to make this examination. The first statistic compares the relative evidence of the two models for fitting the data and resembles the Bayes Factors~(BF). The second statistic quantifies the deviation of inference made using model $\mathbb{M}$ from that made using model $\mathbb{M}_{pl}$ and resembles the $\chi^2$ statistic. We first calculate these two statistics for the \ac{GW} observations and then compare them against statistics that would arise if the true astrophysical distribution were featureless in the chirp mass range 9.5--20$M_\odot$. We create multiple mock catalogs by simulating the observation and measurement process. We assume that the mean of the inferred distribution from the model $\mathbb{M}_{pl}$ is the true astrophysical chirp mass distribution. This choice is data-informed and ensures that the simulated catalogs have a comparable number of observations to the real catalog in the chirp mass range of 9.5--20$M_\odot$.

\begin{center} 
\begin{figure*} 
\centering
\includegraphics[width=0.96\textwidth]{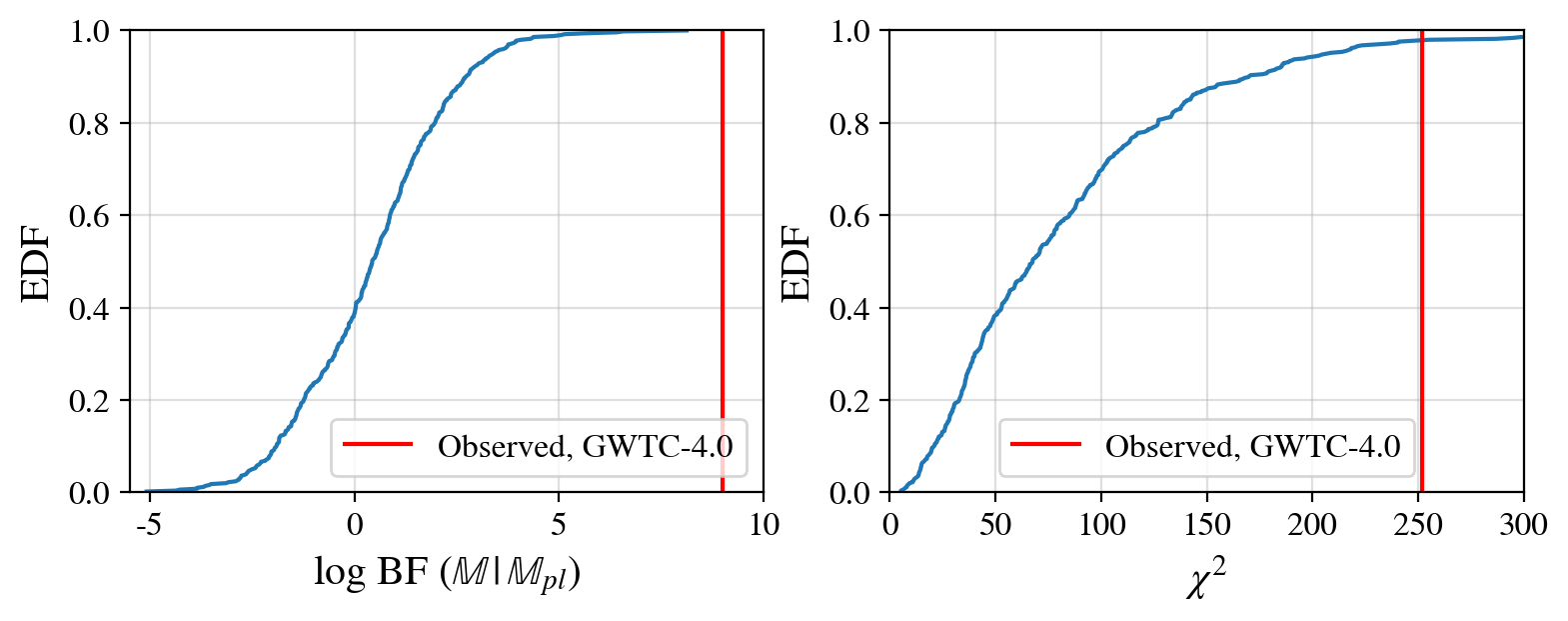} 
\caption{The log Bayes factor (left) and $\chi^2$ values (right) between models $\mathbb{M}$ and $\mathbb{M}_\mathrm{pl}$ for the 500 simulated catalog. The red lines indicate the estimated values for the GW observations using the GWTC-4.0 data.}
\label{fig:conf_peak2} 
\end{figure*} 
\end{center}

Figure~\ref{fig:conf_peak2} compares these two statistics for the \ac{GW} observations against the ones obtained using 500 mock catalogs. BF for the \ac{GW} observations is larger compared to the ones obtained for all the mock catalogs; thereby providing a confidence of greater than 99.5\% in the presence of a feature. The $\chi^2$ value for the \ac{GW} observations is larger than 98\% of those obtained from the mock catalogs, thereby giving 98\% confidence in the presence of a feature. 

Although we estimated the confidence in the second peak using the previous version of the model that directly inferred the chirp mass distribution, we have shown in \ref{apndx:confpeak2} that the chirp mass inferences are very close between the previous and the presented versions of Vamana. 

\section{Inferring Black Hole Mass and Spin Distributions}
\label{apndx:hispin}
Vamana infers the distribution on the component mass plane using truncated Normals. The data products impose a condition $m_2 < m_1$, i.e., the primary mass is always larger than the secondary mass. An alternative approach is to project the data onto the component mass plane, thereby eliminating the need to truncate the multivariate Gaussians~\citep{2025arXiv250602250S}. The model can then be modified to infer the component mass distribution directly, but this also requires inferring the pairing probability between the \acp{BH}, as we need to infer two mass parameters. We have only a limited number of observations, and we fix the mass-ratio distribution and redshift evolution to the standard prior used in parameter estimation~\citep{2019ApJS..241...27A}. Moreover, we only use a diagonal covariance matrix. Therefore, we have just a one-dimensional mixture model to infer the component mass distribution and aligned spin distribution using \ac{BH} mass and aligned spin measurements from \acp{BBH},
%
To obtain the posterior on hyperparameters requires calculating the term~\citep{loredo_2005, 2019MNRAS.486.1086M},
\begin{equation}
    p(\Lambda|d) = \prod_{j=1}^{N_\mathrm{obs}}\left(\sum_i\,w_i\sum_k\mathcal{N}(m_1^{j,k}, m_2^{j,k}, s^{j,k}_{1z}, s^{j,k}_{2z}|\Lambda)\right) / p_\mathrm{PE},
    \label{eq:model3}
\end{equation}where $\Lambda\equiv(\mu_i^{j, k}, \sigma_i^m, \mu_i^\chi, \sigma_i^\chi)$ are the model hyperparameters, index $j$ goes over all all the observation, index $i$ goes over all the mixing components and index $k$ goes over the parameters estimated for an observation obtained using the standard prior, $p_\mathrm{PE}$. As we have fixed the prior on the redshift evolution of the merger rate and mass ratio, the standard prior is $p_\mathrm{PE} \propto (1 + z^{j, k}) ^ 2$ because it is uniform on the detector frame masses. Similarly, spins are assumed to be uniform in magnitude and isotropically distributed; the prior on the aligned component of the spin is $p_\mathrm{PE}\propto -\log(|s^{j,k}_{1z}|) / 2$ and $p_\mathrm{PE}\propto -\log(|s^{j,k}_{2z}|) / 2$. To evaluate Equation~\ref{eq:model3} we just have to calculate terms:
\begin{enumerate}
    \item $\sum_k\mathcal{N}(m_1^{j,k}|\mu_i^m, \sigma_i^m) / (1 + z^{j, k})$,
    \item $\sum_k\mathcal{N}(m_2^{j,k}|\mu_i^m, \sigma_i^m) / (1 + z^{j, k})$,
    \item $\sum_k-2\mathcal{N}(s_{1z}^{j,k}|\mu_i^\chi, \sigma_i^\chi)/\log(|s^{j,k}_{1z}|)$,
    \item $\sum_k-2\mathcal{N}(s_{2z}^{j,k}|\mu_i^\chi, \sigma_i^\chi)/\log(|s^{j,k}_{2z}|)$.
\end{enumerate}
The product of these terms is calculated for each component. A weighted sum is calculated for each observation. The sums of all the observations are multiplied to obtain the final likelihood. When inferring the component mass distribution shown in Figure~\ref{fig:hispin_pm}, we ignore the contribution of any \acp{BH} with a mean secondary mass greater than 40$M_\odot$ when evaluating the likelihood. This restricts the focus to inferring only the first three peaks, as the third peak is located at approximately $30M_\odot$.

The parameters of \acp{BBH} are estimated using multiple waveform families. For the current and past catalog, the most recent release uses {\sc{IMRPHENOMXPHM}}~\citep{2021PhRvD.103j4056P, 2025PhRvD.111j4019C}, {\sc{SEOBNRV(4}} or {\sc{{5)PHM}}}~\citep{{2023PhRvD.108l4035P, 2023PhRvD.108l4037R}} and {\sc{NRSUR7DQ4}}~\citep{2019PhRvR...1c3015V} to estimate the parameters. Due to waveform systematics, the estimated parameters from different waveform families do not always agree, and there is a lack of robust methodology to address them. To infer the \ac{BH} mass and spin distribution from high-spin binaries, we choose parameters estimated using the \sc{SEOBNRV(4} or \sc{{5)PHM}}. The chosen observations are listed in Table~\ref{tab:mean_mass_hspin}.

\begin{deluxetable}{cc|cc|cc}
\tablecaption{Mean primary and secondary BH masses for the high-spin BBHs. These values approximately cluster around 8.5, 16.2, and 30.7 $M_\odot$. \label{tab:mean_mass_hspin}}
\tablewidth{0pt}
\tablehead{
\colhead{Observation} & \colhead{Mean masses ($M_\odot$)} & \colhead{Observation} & \colhead{Mean masses ($M_\odot$)} & \colhead{Observation} & \colhead{Mean masses ($M_\odot$)} 
}
\startdata
GW151226 & 15.5--7.5 & GW170729 & 53.1--33.9 & GW190412 & 30.2--8.5 \\
GW190513 & 35.6--19.5 & GW190517 & 38.3--25.3 & GW190620 & 58.3--36.9 \\
GW190719 & 37.7--21.3 & GW190805 & 47.8--31.3 & GW190828 & 33.1--26.2 \\
GW191103 & 12.6--7.8 & GW230624 & 30.9--16.1 & GW230820 & 63.9--33.2 \\
GW230825 & 48.0--29.2 & GW230928 & 57.0--31.0 & GW231113 & 41.5--27.6 \\
GW231118 & 19.5--11.3 & GW231223 & 47.1--30.5 & GW231230 & 58.0--34.9 \\
GW240107 & 61.3--29.7 & GW241011 & 18.9--6.2 & GW241110 & 16.8--8.3  \\
\enddata
\end{deluxetable}

\bibliography{references}{}
\bibliographystyle{aasjournalv7}



\end{document}